\documentclass[twocol]{ametsoc}
\journal{jas}

\bibpunct{(}{)}{;}{a}{}{,}

\newcommand{\wm}{\,W/m$^2$}
\newcommand{\mwk}{\,mW/m$^2$K}

\title{Entropy production rates of the climate} 

\authors{Goodwin Gibbins\correspondingauthor{Department of Physics, Imperial College London, Prince Consort Rd, Kensington, London SW7 2BW, UK}} 

\affiliation{Imperial College London, UK}

\extraauthor{Joanna D. Haigh}
    \extraaffil{Imperial College London, UK, and }
    \extraaffil{Grantham Institute - Climate Change and the Environment, Imperial College London, UK}

\email{r.gibbins15@imperial.ac.uk}

\abstract{There is ongoing interest in the global entropy production rate as a climate diagnostic and predictor, but progress has been limited by ambiguities in its definition; different conceptual boundaries of the climate system give rise to different internal production rates. Three viable options are described, estimated and investigated here, two of which -- the material and the total radiative (here `planetary') entropy production rates -- are well-established and a third which has only recently been considered but appears very promising. This new option is labelled the `transfer' entropy production rate and includes all irreversible processes that transfer heat within the climate, radiative \textit{and} material, but not those involved in the exchange of radiation with space. Estimates in three model climates put the material rate in the range $27$-$48$\mwk, the transfer rate $67$-$76$\mwk, and the planetary rate $1279$-$1312$\mwk. 
\\
The climate-relevance of each rate is probed by calculating their responses to climate changes in a simple radiative-convective model. An increased greenhouse effect causes a significant increase in the material and transfer entropy production rates but has no direct impact on the planetary rate. When the same surface temperature increase is forced by changing the albedo instead, the material and transfer entropy production rates increase less dramatically and the planetary rate also registers an increase.  This is pertinent to solar radiation management as it demonstrates the difficulty of reversing greenhouse gas-mediated climate changes by albedo alterations. It is argued that the transfer perspective has particular significance in the climate system and warrants increased prominence.
}

\begin{document}

\maketitle


\section{Introduction}

\subsection{Motivation} 

The climate is, fundamentally, an entropy-producing system. The movement of energy from warmer regions, where it is supplied to the climate, to cooler regions, where it leaves, is an inevitable consequence of the second law of thermodynamics and drives the motion and activity of the climate. The energy transfers are mediated by a myriad of irreversible processes, for example wind, rain and radiation. Each process produces entropy, which must be exported from the system in order to maintain a steady state. The export is by radiation; the supply of low-entropy solar radiation and loss of high-entropy outgoing thermal radiation has the net effect of carrying entropy away from the system and maintaining temperature gradients. Our climate system exists in this balance.

Although entropy in the climate system has been explored in the literature for more than four decades, its utility in studying the climate has not been established. A major limitation has been the difficulty in pinning down even the concept of a global entropy production rate, which has been challenging both because an intuitive understanding of the entropics of the climate is difficult to develop and because there are multiple candidate climate-relevant global entropy production rates that can be defined for different notions of the system's extent and boundaries, as recently underlined by \cite{Bannon2015}. 

Our first purpose here is to clarify further the different definitions of global entropy production rates. The two main perspectives that have gained traction in the literature are one that considers the entropy production due to all radiative and non-radiative irreversible processes (here labelled \textit{planetary}) and one that restricts itself to the non-radiative irreversible processes only (labelled \textit{material}). We advance a third, labelled the \textit{transfer} entropy production rate, which accounts for the entropy produced by all processes that transfer heat within the climate system -- radiative and non-radiative -- but not that associated with the thermalization and scattering of incoming solar radiation or with the emission of outgoing thermal radiation to space. 

A summary of the historical evolution of the material and planetary perspectives is given in Section \ref{sec:context}, leading to an argument for the addition of the transfer perspective. Definitions of the material, transfer and planetary entropy production rates are then offered in Section \ref{sec:EPRs} using consistent notation, underlining their differences and interpretation. The measurement of all three entropy production rates is demonstrated in Section \ref{sec:values} in three climate models: an energy balance model, an analytic radiative-convective model and an observationally-based standard atmospheric column.

Our second purpose is to explore the behavior and possible uses of these global entropy production rates in quantifying and understanding the climate. We approach this in Sections \ref{sec:RCM}a and \ref{sec:RCM}b by demonstrating how each entropy production rate responds to greenhouse gas concentration and albedo changes in the analytic radiative-convective model. This leads to further insight into the physical significance of the three entropy production rate perspectives in Section \ref{sec:RCM}c.

\subsection{Using entropy production rates}
What is an entropy production rate? Returning to the second law of thermodynamics, recall that the entropy of the universe increases due to each irreversible process. If these processes occur inside an isolated, closed system, that increase must be reflected in the entropy ($S$) of the system and they are said to have `produced entropy' inside the system at a rate $\Sigma$: 
\begin{equation}
\frac{dS_{closed}}{dt} = \Sigma \ge 0
\end{equation}

This can be extended to systems that are open to certain cross-boundary fluxes but are steady, such that their properties do not change in time and there is no net storage within the system. These driven non-equilibrium systems are useful first-order approximations for the Earth.  

For a conserved quantity, such as energy $U$, there is a balance between the fluxes, $F$, into and out of a steady system:
\begin{equation}
\frac{dU_{open,\ steady}}{dt} = 0 = F_{in} - F_{out}
\end{equation}
This feature is already well-mobilized for simplifying, constraining and explaining aspects of the climate system, as in energy balance models and the concept of radiative forcing.

However, as entropy can be created but not destroyed within the system, a different kind of balance emerges: the cross-boundary flow must carry a net outwards flux\footnote{Note that $J$ is the positive rate of entropy change experienced \textit{by the system} due to the cross-boundary flux. If the import or export processes are themselves irreversible, the synchronous entropy change perceived by the surroundings due to the same cross-boundary flow will be different to that perceived by the system.} of entropy ($J_{out} - J_{in}$), which equals the total internal production rate, $\Sigma$:
\begin{equation}
\frac{dS_{open,\ steady}}{dt} = 0 = J_{in} - J_{out} +  \Sigma.
\label{eq:balance}
\end{equation}
This entropy production rate can be identified \citep{Peixoto1991}  `directly' by summing the entropy increases ($\sigma_i$) due to all the irreversible processes that occur within the system:
\begin{equation}
\Sigma = \sum_{i} \sigma_{i}.
\end{equation}
or `indirectly' as the difference of the cross-boundary entropy fluxes:
\begin{equation}
\Sigma = J_{out} - J_{in}
\end{equation}

This is a potentially useful constraint: although entropy is produced across a wide variety of processes within the system, in steady state it must be exactly exported by the cross-boundary fluxes. Note that there is a distinction made here between a flux, which changes the entropy of the system by moving energy across the boundary, and a production, in which movement of energy within a system increases its entropy. 

Heat delivered at rate $F$ to a region at temperature $T$ increases the entropy there at a rate $F/T$, which is identified as an entropy flux into the system. This can be generalized to give a temperature for each paired entropy and energy flux, $T:= F/J$, which represents an average quality of the energy at the point of entry to or exit from the system. For a system with a steady flow of energy $F$ through it, the inflow and outflow temperatures, $T_{in} = F/J_{in}$ and $T_{out} = F/J_{out}$, are sometimes combined into an `efficiency' $\eta = (T_{in} - T_{out})/T_{in}$, as in the Carnot efficiency for the maximum work per unit heat input that can be extracted by a reversible engine operating between two fixed temperatures. In our case, however, the value does not represent work extracted (as there is no mechanism for any extraction), but instead can be interpreted as a summary metric of the irreversibility of the system or the ``lost work'' (e.g. in this context, \citet{Bannon2015}). If heat flows at a rate $F$ from $T_{hot}$ to $T_{cold}$, the entropy production will be exactly $\sigma = F \left(1/T_{cold} - 1/T_{hot} \right)$, provided the mechanism that mediates the transfer is returned to its original configuration. The entropy production rate of the system can then be related to the temperatures, efficiency and energy flow by $\Sigma = F (1/T_{out} - 1/T_{in}) =  F (\eta /T_{out})$.

%
%
%
\section{Context}
\label{sec:context}

\subsection{History}
The idea that an entropy production rate might hold predictive significance for the climate was initially ``stumbled upon'' by Garth Paltridge in the 1970s (his words, \cite{Paltridge2005} describing \cite{Paltridge1975}). Motivated by the idea that the complexity of the climate might lead to emergent control by some extremization principle, he conducted a broad search for climate variables whose maximization or minimization in a simple 1D energy balance model returned realistic meridional heat flows. The most promising and physically interpretable variable he constructed was entropy-like: the net shortwave (SW) and longwave (LW) energy fluxes divided by a representative local temperature, $(F_{SW} - F_{LW}) / T$, summed across the zonal grid cells of his model. 

This was the observation that sparked the field of climate-entropy research. In Paltridge's zonally-averaged model, the difference in radiative heating was interpreted as the meridional heat transfer rate, as further investigated by \cite{Paltridge1978, Nicolis1980, Grassl1981, Wyant1988}, among others. 

\citet{Essex1984i}, however, argued Paltridge's  characterization of the Earth's entropy production rate was fundamentally ``incorrect'' as it had failed to account for entropy production in the radiation field. In doing so, Essex introduced the quantity labelled the \textit{planetary entropy production rate}: ``the entire entropy production of a steady state climate is contained in the difference of the entropy of the outgoing terrestrial radiation from the entropy of the incoming solar radiation''. This planetary rate has been further explored by, e.g. \cite{Lesins1990, Stephens1993, Pelkowski1994, Li1994v, Li1994h, Wu2010}. 

This insight into the role of radiation in producing entropy instigated the division of the climate into non-radiative (material) and radiative sub-systems and the separate tabulations of the entropy produced in each \citep{Essex1987, Goody1996, Goody2000}. Paltridge's meridional view was mapped in higher-dimensional models to the \textit{material entropy production rate}, which includes contributions from horizontal and vertical sensible and latent heating (as established by \cite{Pujol1999}) as well as all other non-radiative processes.

By virtue of its connection with moving, tangible matter, the material perspective has been preferred for application of a maximization principle, both in experimental and theoretical studies (e.g.\ \cite{Ozawa1997, Ozawa2003, Dewar2003, Fraedrich2008, Labarre2019}). The contributions due to component processes have also been considered separately (e.g.\ \cite{Pauluis2002_1, Pauluis2002_2, Volk2010, Lembo2019}) and the response to changing climate conditions studied \citep{Singh2016, Bannon2017}. 

%
%
%
\subsection{Why add the transfer perspective?}

The key insight that supports the transfer entropy production rate is that radiative processes can be further categorized according to the roles they play within the climate. This is not a new perspective - \cite{Green1967} argues for it explicitly - but it has not yet been discussed in detail in the entropy production literature.

Energy is supplied to and exported from the planet by radiation; this \textit{external radiation} on average heats warm places and cools cold places, continually driving temperature differences in the system. By contrast, \textit{internal radiation}, which is emitted and absorbed by the material of the Earth system, gives a net transfer of heat down-gradient, pulling the system towards thermal equilibrium. It is almost coincidental that radiation occurs in both of these roles. Internal radiation is an inherent feature of a translucent, warm atmosphere and would occur even if the system were driven by a non-radiative heat source and sink. It is not fundamentally different nor necessarily distinguishable from material heat transfer processes such as conduction. While external radiation determines where and how much heating and cooling drives the system, internal radiation acts in parallel with material processes to transfer energy between where the external radiation delivers it to and takes it from. 

This similarity in function between internal radiation and material processes suggests that they might be best considered together, as in the transfer entropy production rate. This concept of global entropy production has appeared only occasionally in the literature, and has not yet been thoroughly explored. In \cite{Bannon2015}, one of the material entropy production rates discussed (his MS3) includes the internal radiative heating processes and so is the transfer rate discussed here. In \cite{Bannon2017}, the transfer rate is estimated as an upper bound for the material production rate. Most recently, \cite{Kato2020} presents a study in which the transfer entropy production rate is estimated in Bannon's simple model and from observational data. Here we explore all three entropy production rates and their definitions, estimates and responses to climate change, in order to elucidate the significance of the transfer entropy production rate concept.

%
%
%
\section{Definitions of entropy production rates}
\label{sec:EPRs}

The multiplicity of entropy production rates for the Earth's climate arises because there is not a self-evident boundary of the system with respect to radiation. Different perspectives can give different extents of the system, not in terms of physical space (all extend from the lithosphere to the upper atmosphere) but in terms of how or when radiation or heat crosses into and out of the system.

The interaction of radiation with matter naturally suggests the three perspectives on the climate system that were introduced above and can be succinctly described by the nature of the energy fluxes that cross their boundaries:

\begin{description}
\item[ Planetary:] photons carry entropy and energy into and out of the system as they cross a control volume surface beyond the top of the atmosphere.
\item[Transfer:] energy enters the system when it is first absorbed by matter on its way from the sun and leaves when it is last emitted on its way to space.
\item[Material:] every absorption or emission of radiation is a crossing of energy into or out of the material system. Energy is only in the system when it is in matter, not as photons. 
\end{description}

These systems are nested (listed here from large to small), have different cross-boundary entropy fluxes and include different sets of irreversible processes.  Thus each measures a different global entropy production rate, although they all, in some sense, self-consistently describe `the climate'. This section explains their definitions, distinctions, and interpretations; the question remaining for scientists trying to use entropy production rates as a predictive or diagnostic variable will be their physical relevance. 

\begin{figure*}[t]
\begin{center}
  \noindent\includegraphics[width=.7\textwidth, angle=0]{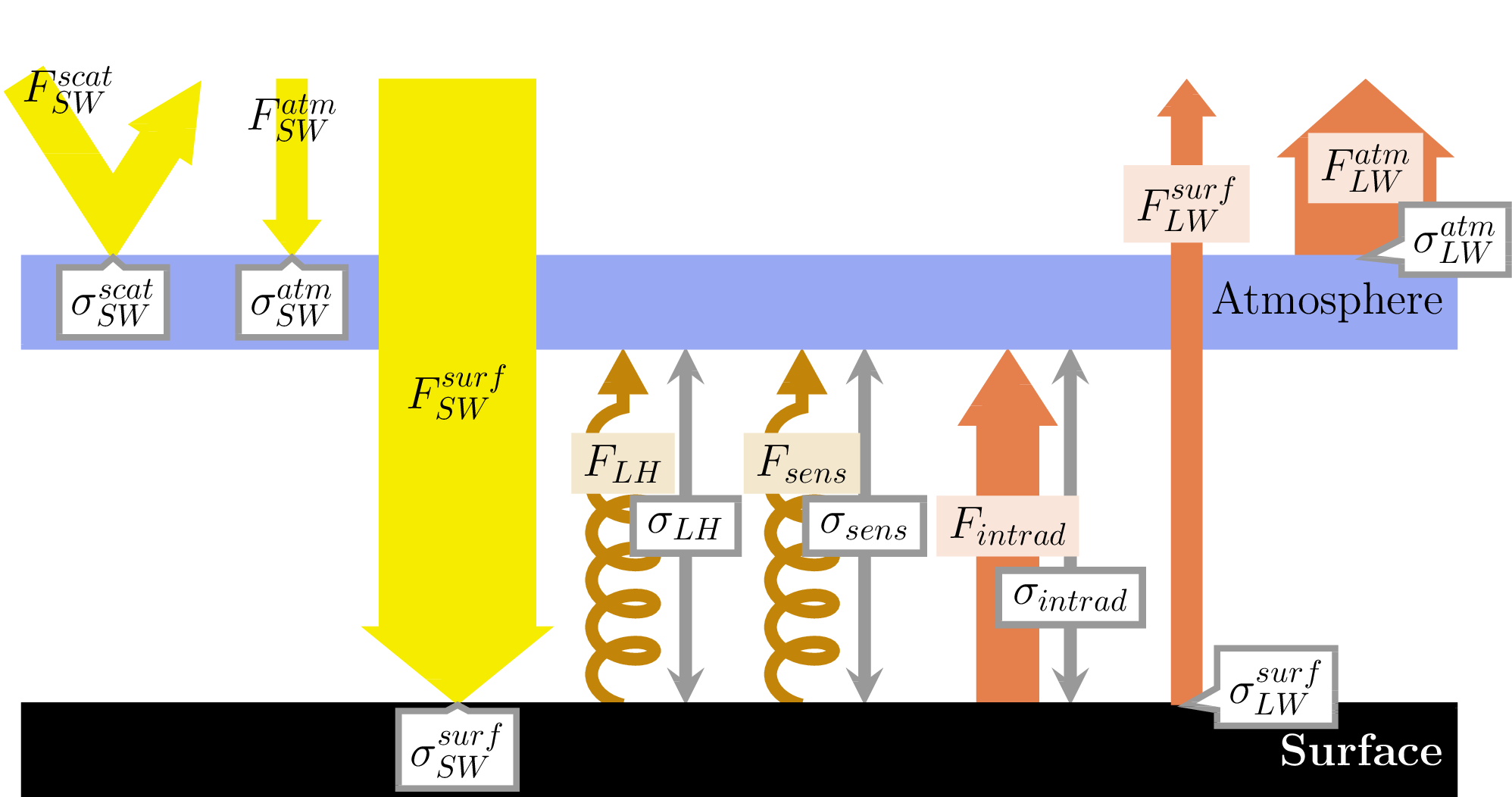}\\
  \caption{ A schematic of a simple two-layer zero-dimensional energy balance model with energy fluxes labelled as colored arrows $F$ and the resulting entropy productions boxed, $\sigma$. The energy supplied to the system from the surroundings is shown entering or exiting from the top of the diagram. Three internal energy exchanges are shown: two material processes -- latent and sensible heat fluxes -- and internal radiation. Scattering from the surface has not been shown but is implied. This diagram mirrors the energy balance study by \cite{Wild2014} and is constructed to match \cite{Bannon2015}.}
\label{fig:ebm}
\end{center}
\end{figure*}

An energy balance model (EBM) adapted from \cite{Bannon2015} and shown graphically in Figure \ref{fig:ebm} is used to demonstrate which processes are included in each entropy production rate. It can be solved analytically (see Appendix A) to give the surface and atmosphere temperatures as a function of albedo, emissivities and latent and sensible heat flux.

The model broadly echoes the global energy budget diagrams of \cite{Wild2014}, but with a non-standard rearrangement of the radiative energy fluxes to separate the internal radiation (net flux between surface and atmosphere) from the external radiation (fluxes that leave to or enter from the surroundings), reminiscent of the Net Exchange Formulation discussed by \cite{Herbert2011gig}. The two material terms - latent and sensible heat transfer - should be taken as placeholders for the full range of non-radiative energy transfer mechanisms, including via creation and dissipation of kinetic energy in the general circulation. Each down-gradient energy transfer causes an entropy production, which are labelled in rectangular boxes. Their values are calculated in Appendix B.

For simplicity, heat from the planetary interior and the irreversibility of life are not explicitly treated here, but the definitions could be adapted to feature them. Note also that the entropy production rates and fluxes defined are for the whole system, globally and not locally, although they are quoted in per-area units (\mwk). The symbol $\sigma$ refers to an entropy production rate due to a particular type of irreversible process, while $\Sigma$ is the aggregated global value for the system in question.

%
%
%
\subsection{The planetary entropy production rate}

The planetary entropy production rate can be interpreted as the entropy change of the universe due to the existence of a planet interrupting and thermalizing the solar photons: it is the difference in entropy of the photons incident on the Earth compared to those scattered and radiated away from it. The system is defined via a control volume that surrounds the planet and includes the entropy production due to all irreversible processes that happen within it, radiative or otherwise (see CV1 in \cite{Bannon2015}).  

Defined directly for the energy balance model of Figure \ref{fig:ebm}, the planetary rate is the sum of all the entropy budget items, radiative and non-radiative:
\begin{equation}
\Sigma_{planet} = \sigma^{scat}_{SW} + \sigma^{atm}_{SW} + \sigma^{surf}_{SW} +  \sigma_{LH} + \sigma_{sens} + \sigma_{intrad} + \sigma^{surf}_{LW} + \sigma^{atm}_{LW}
\label{eq:PEPR_dir}
\end{equation} 

To define the production indirectly, the entropy content ($L_\nu$) associated with the flow of photons of frequency $\nu$ (units of $1/$s) as they cross over the boundary of the control volume can be calculated as a non-linear function of their spectral intensity $I_\nu$: 
\begin{equation}
L_\nu = (1+y)\ln (1+y) - y \ln (y)
\label{eq:spect}
\end{equation}
where
\begin{equation}
y = \frac{c^2}{n_0 h \nu^3} I_\nu.
\end{equation}
Here $c$ is the speed of light and the number of polarization states $n_0 = 2$.  Equation \ref{eq:spect} can be derived by analysis of photons as a boson gas (\cite{Planck1914}, further developed by \cite{Rosen1954}) or  from the relationship $dL_\nu = dI_\nu / T$ where $I_\nu = B_\nu(T)$ is the spectral Planck function intensity, which is integrated from zero to the relevant intensity \citep{Ore1955}. 

The entropy flux is then calculated by integrating this over all wavelengths, in the inbound or outbound hemispheric directions and averaged over the surface area of the planet:
\begin{equation}
J^{in/out}_{planet} =\frac1A \int dA \int d\nu \int_{\text{hemi}} \cos\theta d\Omega\ L_\nu.
\label{eq:Jplanet}
\end{equation}
 
This entropy flux in radiation simplifies in the case of a black body to a simple dependence on source temperature $T_{source}$, as explained by \cite{Planck1914} and \cite{Essex1984i} and  demonstrated in the appendix to \cite{Wu2010}:
\begin{equation}
J_{BB}= \frac43 \frac{F}{T_{source}} = \frac43 \sigma T_{source}^3.
\end{equation}
The factor of $4/3$ in the emitted entropy of the radiation compared to the entropy change of the emitting material ($J_{mat} = F/T_{source}$) accounts for the irreversibility of radiating into a vacuum \citep{Feistel2011e}. From the planetary perspective, it is this larger entropy flux in the radiation that crosses the system's boundary.

Scattered solar photons carry much less entropy than those thermalized and re-emitted by the planet and so the albedo of the Earth is a key determinate of $\Sigma_{planet}$. The planetary perspective has been further subdivided in some entropy studies to give related variables, for instance by excluding the production from the fraction of the solar radiation that is scattered (CV2 in \citet{Bannon2015}) or by focusing on only the production which occurs within the atmosphere \citep{Peixoto1991}. However, we would argue that the definition as given here is more fundamental.

%
%
%
\subsection{The material entropy production rate}

The material entropy production rate avoids dependence on spectral properties of radiation and the solar temperature by excluding \textit{all} radiative processes from the entropy tally. In this view, the system is exclusively the matter. The photon gas permeating the atmosphere is considered part of the surroundings and radiative heating and cooling supply the cross-boundary fluxes of energy \citep{Bannon2015}. The local temperature distribution of the matter is unchanging at steady state and so the net heating of each parcel by material processes must be balanced by radiative cooling to space or within the climate system, and vice versa \citep{Essex1987, Goody2000}.  

The view that motivates this approach is that these material (or `molecular') processes -- such as phase changes, sensible heating, friction and diffusion -- are the ones of interest in the dynamics of the weather and other tangible aspects of the climate. The material entropy production has been linked to the kinetic energy conversion in the Lorenz Energy Cycle (for example \cite{Lucarini2011}) but includes processes that are not related to motion as well, such as diffusion. In a moist atmosphere, the material entropy production is dominated by contributions from the hydrological cycle \citep{Pauluis2002_1}.

In our EBM (Figure \ref{fig:ebm}), the material entropy production rate can be directly specified as:
\begin{equation}
\Sigma_{mat} = \sigma_{LH} + \sigma_{sens}.
\end{equation}

The net entropy flux into and out of the material system due to all radiative heating defines it indirectly:
\begin{equation}
\Sigma_{mat} = \int dV\ \frac{-\dot{Q}_{rad}(\mathbf{x})}{T_{mat}(\mathbf{x})} = J^{out}_{mat} - J^{in}_{mat}
\label{eq:mat}
\end{equation}
where $\dot{Q}_{rad}$ is the local radiative heating rate (in W/m$^3$) and $T_{mat}$ is the temperature of the material where the heating or cooling occurs. There are multiple options for partitioning this net radiative heating into a $J^{in}_{mat}$ and $J^{out}_{mat}$. The two approaches described by \citet{Bannon2015} are to consider all absorption of radiation separately from emission (his MS1), or alternatively to separate the solar SW heating from the LW radiative heating and cooling (his MS2). Here we follow \citet{Lucarini2009} in distinguishing regions of net positive $Q_{rad}$ from those with net negative, which takes advantage of the fact (discussed in \cite{Goody2000}) that, in steady state, the net local radiative heating must be balanced exactly by non-radiative cooling (and vice versa) and so the temperatures and energy fluxes calculated in this way will closely reflect those experienced by the material processes. These approaches lead to different analyses of the energy and entropy fluxes, entropic temperatures and efficiency of the material system, but the same production rate. 

Paltridge's original meridional heat transport entropy production rate can be interpreted as the horizontal component of the material entropy production rate, which has been estimated to be approximately $15\%$ of the total \citep{Pascale2012}. 

%
%
%
\subsection{The transfer entropy production rate}
The transfer system includes matter and the internal radiation that travels between matter within the climate system but not the external radiation before it has interacted with the matter or after it has been emitted for the last time, which is considered part of the surroundings. 

The observation that motivates the transfer perspective -- that internal radiation plays a role in the climate that is parallel to the material processes -- is particularly evident in the entropy productions. Since internal radiation is emitted and re-absorbed again within the system, only the heat transfer it causes, and not its entropy while a photon gas, is relevant to the global entropy production budget.  Therefore, the entropy production due to internal radiation, $\sigma_{intrad}$, takes the same form as the entropy production due to the material processes in our two-layer model: $F (1/T_{atm} - 1/T_{surf})$, where $F$ is the rate of energy transport. By contrast, the five entropy productions relating to external radiation that are excluded directly depend on the details of the radiation spectra.

Defined directly for our EBM (in Figure \ref{fig:ebm}):
\begin{equation}
\Sigma_{tran} =  \sigma_{LH} + \sigma_{sens} + \sigma_{intrad}
\end{equation}
which makes it intermediate in magnitude:
\begin{align}
\Sigma_{tran} & =  \Sigma_{mat} + \sigma_{intrad}  \\&=  \Sigma_{planet}  - (\sigma^{scat}_{SW} + \sigma^{atm}_{SW} + \sigma^{surf}_{SW} +   \sigma^{surf}_{LW} + \sigma^{atm}_{LW})
\end{align}
as it includes contributions from the internal radiation, which are not in the material entropy production rate, and excludes those from external radiation, which are included in the planetary rate.

In the transfer system,  the cross-boundary fluxes are the heating or cooling of material upon absorption or emission of external radiation and so the production can be written:

\begin{equation}
\Sigma_{tran}  = \int dV\ \frac{-\dot{Q}_{\text{ext rad}}(\mathbf{x})}{T_{mat}(\mathbf{x})} = J^{out}_{tran} - J^{in}_{tran}
\label{eq:tran}
\end{equation}

The partitioning of the external radiative heating into $J^{in}_{tran}$ and $J^{out}_{tran}$ components is, like in the material case, a further definitional choice. One option would be to separate areas of net positive external radiative heating from areas of net negative, however we take the more straightforward approach of separating the absorption of solar radiation from the emission of long-wave radiation to space. 

Then the incoming entropy flux is due to the absorption of solar radiation:
\begin{equation}
J^{in}_{tran} = \int dV\ \frac{\dot{Q}_{sw}(\mathbf{x})}{T_{mat}(\mathbf{x})}
\label{eq:Jtransin}
\end{equation}
where $\dot{Q}_{sw}$ is the heating rate due to SW radiation and $T_{mat}$ is the temperature of the absorbing material. 

The entropy flux out of the system is due to the LW emission of radiation that is not reabsorbed within the system. This is the cooling to space, $\dot{Q}_{cts}$, which can be calculated by radiative transfer models if the optical depth and temperature ($T$) are known \citep{Rodgers1966, Wallace2006}: 
\begin{equation}
\dot{Q}_{cts}(z) = - \pi \int d\nu\ B_\nu(T) \frac{d\mathcal{T}_\nu(z,\infty) }{dz} 
\label{eq:cts}
\end{equation}
where $\mathcal{T}_\nu(z,\infty) $ is the transmittance between $z$ and the top of the atmosphere. Then the outwards entropy flux for the transfer system becomes:
\begin{equation}
J^{out}_{tran} = - \int dV\ \frac{\dot{Q}_{cts}(\mathbf{x})}{T_{mat}(\mathbf{x})}
\label{eq:Jtransout}
\end{equation}
 Note that the cooling to space is necessarily negative and its sum is exactly the outgoing longwave energy flux leaving the planet. It is the temperature from which cooling to space occurs, along with the spectral properties of the radiatively active gases, that determines the shape of the outgoing emission spectra. The average cooling-to-space temperature will generally be close to the emission temperature of the planet.  
 
In the horizontal, there is negligible net heat transfer by internal radiation and so the horizontal components of the transfer and material entropy production rates converge. Therefore, the meridional heat transfer entropy production rate of Paltridge can equally be identified as the horizontal component of $\Sigma_{tran}$.

%
%
%
\section{Estimates of the entropy production rates}
\label{sec:values}

We now use these definitions to estimate each entropy production rate and related variables in three model climates of increasing complexity. 

For the EBM of \cite{Bannon2015}, the entropy production due to each process can be calculated separately (Appendix B) and the entropy production rates directly summed. The indirect approach, which focuses on radiation fluxes, can equally be applied, as explored in Appendix C. The estimated entropy production rates are listed alongside the fluxes, implied influx and outflux temperatures and resulting efficiencies in the first section of Table \ref{tab:values}. 

\begin{table*}[h!]
\begin{center}
\vspace{.4em}
\begin{tabular}{l || ccc |  ccc | ccc }
\multicolumn {1}{c}{ } & \multicolumn{3}{c}{\textit{EBM}} &   \multicolumn{3}{c}{\textit{RCM}} &  \multicolumn{3}{c}{\textit{Std Atmos}}\\
\hline\hline
 					& Planet	& Tran	&   Mat   	&  Planet 	&   Tran	& Mat 	& Planet	 & Tran  	& Mat	\\
\hline
 $F^{in} = F^{out}$ (W/m$^2$) & 341& 239	& 85	   	&    343	&	240	&	112	&   346	&    259	&  96		\\ \hline
 $J^{in}$ (mW/m$^2$K) 	& 79 		& 869 	& 306  	&	79	&   857	&	389	&   79	&   929	&  334	\\
 $J^{out}$ (mW/m$^2$K)	& 1358 	& 936 	& 336  	&  1365	&  933	&	416	& 1391	&   999	&  382 \\ \hline
 $T^{in}$ (K) 			& 4334 	& 275 	& 278 	&    4334	&	280	&	288	&  4364	&   280	&  288 \\
 $T^{out}$ (K) 			& 251	& 255 	& 253 	&     251	&	257	&	269	&  248	&   260	&  252 \\ \hline
 $\eta$ (\%)	 		& 94.2	& 7.2 	& 8.6 	&   	94.2	&	8.1	&	6.5	&  94.7	&   7.0 	&  12.5  \\ \hline

 $\Sigma\ $ (mW/m$^2$K)& 1279 	& 67 		& 30	 	&  1286	&	76	&	27	&  1312	&	70	& 48 \\
\hline
\end{tabular}

\end{center}
\caption{The energy fluxes ($F$), associated entropy fluxes ($J$), temperatures ($T = F/J$), efficiencies ($\eta = (T_{in}-T_{out}) / T_{in}$) and production rates ($\Sigma$) for the energy balance model, radiative-convective model and for a clear-sky standard atmospheric column from each of the three system perspectives: planetary, transfer and material.}
\label{tab:values}
\end{table*}

To increase the fidelity with which the climate is represented while maintaining the possibility of analytic solutions, we use the analytic radiative-convective model of \cite{Tolento2019}, which was originally designed for flexibility in capturing a wide range of planetary climates. It approximates the atmosphere as a radiatively gray gas with a convective tropospheric region and a stratosphere in radiative balance with two channels of shortwave absorption, and is described in more detail in Appendix D. It is powerful because analytic expressions for the temperatures and radiative fluxes in the model allow for exact calculation of the entropy production rates via the indirect definitions. The entropy production rate results are quoted in the second section of Table \ref{tab:values}.

As the complexity of the climate model increases, the value of the indirect method becomes more apparent: it is much simpler to characterize the cross-boundary radiation or radiative heating than to quantify the contribution from every irreversible process within the system. Some of the necessary radiative information (such as shortwave and longwave heating rates) is supplied as standard in some reanalysis products, but for calculation of the detailed spectra for the planetary rate or the cooling to space for the transfer rate offline radiative transfer calculations are needed. We explore this approach in the clear-sky standard atmospheric profile of \citet{Anderson1986}, using the radiative transfer software Libradtran \citep{Libradtran2} to recover spectrally and vertically resolved optical depths, irradiance and heating rates.  The surface is treated as a black body with temperature $288.15$\,K and the flux from the overhead sun is scaled such that the incoming and outgoing energy fluxes balance. A standard aerosol profile is used \citep{Shettle1989} and the surface albedo is set to $0.3$ in the shortwave. The single column is interpreted as representing a zonally- and meridionally-symmetric steady planet (with no storage of energy or entropy), but a similar approach could be applied to spatially-varying atmospheric columns. 

In the indirect method, the entropy production rate is calculated as the difference between incoming and outgoing entropy fluxes. For the material system, these can be calculated from net radiative heating rates via equation \ref{eq:mat}. The transfer rate outgoing flux (equation \ref{eq:Jtransout}) first requires calculation of cooling to space first via equation \ref{eq:cts}, while the incoming flux is calculated from the net solar heating rate as in equation \ref{eq:Jtransin}. The planetary entropy fluxes are accessed by application of equations \ref{eq:spect} and \ref{eq:Jplanet} to the radiation spectra. The resulting values are shown in the third section of Table \ref{tab:values}.

These estimates suggest $\Sigma_{mat} \approx 27$-$48$\mwk,  $\Sigma_{tran} \approx 67$-$76$\mwk\ and $\Sigma_{planet} \approx 1279$-$1312$\mwk, which are broadly in agreement with the values calculated in the literature. The material rate has been estimated identically by \cite{Bannon2015} at $30$\mwk\ in the EBM, and in more realistic models by \cite{Pascale2011c} at $\approx 50$\mwk, by \cite{Kato2020} at $49$\mwk\ and by \cite{Lembo2019} in the range $38.7-43.4$\mwk\ (outlier neglected) by their direct method. The planetary rate is also confirmed in the EBM by \cite{Bannon2015} (his CV1), and corroborated by estimates of $1272-1284$\mwk\ by \cite{Wu2010}. The transfer rate calculated by \cite{Kato2020} is $76$\mwk, and is estimated in \cite{Bannon2017} at $68$\mwk. The transfer rate discussed here differs slightly from the closest analogue in \cite{Bannon2015} in the outflow temperature; for Bannon's MS3 the atmospheric temperature ($253$\,K) is used rather than the cooling-to-space weighted average ($255$\,K).

The thermalization of solar radiation is the largest source of entropy production in the planetary view, accounting for more than $60\%$ of the total (see Appendix B). The transfer entropy production rate sums the contributions from the material processes and internal radiative heat transfer, which are of similar orders of magnitude: in the EBM, $\sigma_{LH} + \sigma_{sens}= 30.5$\mwk\ while $\sigma_{intrad} = 36.5$\mwk. The flux $F_{planet}$ is the total incident solar flux, $F_{tran}$ is the fraction absorbed, and $F_{mat}$ focuses on the fraction of energy transferred by material processes. The material $T^{in}_{mat}$ is the surface temperature, as that is where there is net radiative heat input, while the $T^{out}_{mat} $ is an atmospheric average. The transfer $T^{in}_{tran}$ reflects the average temperature of solar absorption and the $T^{out}_{tran}$ is the average temperature of thermal emission to space, which is approximately the effective emission temperature, $(F_{tran}/\sigma)^{1/4}$. The planetary $T^{in}_{planet}$ comes from the solar temperature (scaled because of the $4/3$ factor associated with the emission of radiation) and the $T^{out}_{planet}$ reflects an average of the scaled emission temperature and the high temperature of the scattered solar photons. The efficiencies capture the magnitude of these temperature differences.

%
%
%
\section{Sensitivity of EPRs to climate changes}
\label{sec:RCM}

With three different global entropy production rates in hand, a natural next question is how they respond to changes in the climate state. This offers insight about their interpretation as well as how they might be leveraged as diagnostic variables.

%
%
%
\subsection{Experimental set-up}
The analytic radiative-convective model (described in Appendix D) is well-suited for studying the effect of forcing in the longwave and shortwave on the entropy of the atmosphere, as convection, internal radiation, stratospheric absorption of shortwave radiation and thermal emission of radiation are all handled explicitly and in a simplified manner: convection by fixing a prescribed lapse rate and radiation by assuming a radiatively gray gas in one longwave and two shortwave channels. This allows for climate changes to be treated independently -- with cloud and lapse rate feedbacks suppressed -- to isolate the first-order responses in the vertical.  

The greenhouse effect is simulated by increasing the thermal optical depth by 25\% while keeping the absorption profile of solar radiation fixed, which results in an increase in the surface temperature by $5.5$\,K from $287.9$\,K to $293.4$\,K. The effect of an equivalent temperature increase by means of a top-of-atmosphere albedo decrease (absorptivity increase) from $0.30$ to $0.25$ is compared.

Although the surface temperature change is the same in these cases, other aspects of the climates differ. This has implications for the ability of uniform solar radiation management-type geoengineering to restore a climate with a heightened greenhouse effect to its pre-industrial state. To investigate this, a fourth case is modeled, where to counteract the elevated greenhouse effect, the albedo is also increased to $0.35$, restoring the surface temperature.

%
%
%
\subsection{Results and discussion}

\begin{figure*}[t]
\begin{center}
  \includegraphics[width=.8\textwidth, angle=0]{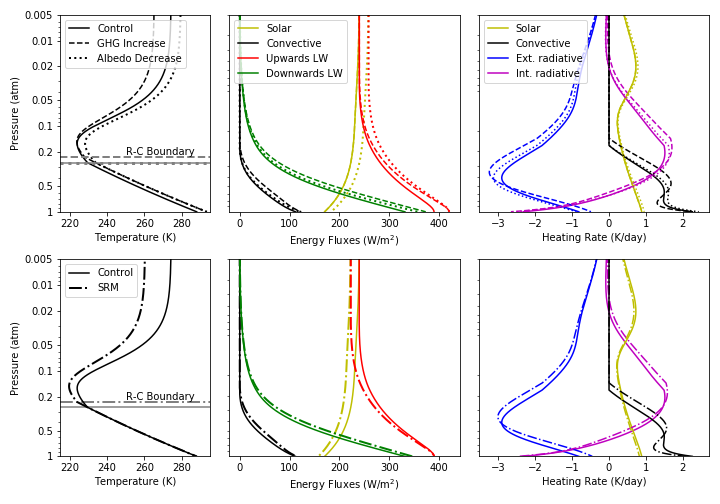}\\
  \caption{The temperature (first column), energy fluxes (second column) and heating rate (third column) profiles derived from the analytic radiative-convective model in a range of climate states. The solid lines show the control case (representing a pre-industrial scenario) with parameters from \cite{Tolento2019} chosen to match observation, with a surface temperature of $287.9$\,K. A $25$\% increase in the longwave atmospheric absorption results in a temperature increase to $293.4$\,K (shown in dashed lines), which is also attained by changing the shortwave absorptivity of the planet via a decrease in the top of atmosphere albedo from $0.30$ to $0.25$ (dotted lines). In the lower panels, the surface temperature is restored by increasing the albedo to $0.35$ to balance the $25$\% increase in greenhouse effect, as in a globally-uniform solar radiation management intervention. In the second column, the longwave energy flux is separated into upwelling and downwelling components, while in the third column the longwave heating rate is separated into that due to cooling to space (external radiative heating, blue) and that due to internal radiative heat transfer within the atmosphere and with the surface (magenta).}
\label{fig:RCM}
\end{center}
\end{figure*}

The atmospheric profiles under these four climate conditions are represented in Figure \ref{fig:RCM}, where the first column shows the temperature profiles, the second the vertical energy fluxes and the third the resulting heating rates. Identical surface temperatures give rise to identical tropospheric temperature profiles because of the prescribed lapse rate, as in the greenhouse gas and increased solar absorption cases (upper panels) and pre-industrial and solar radiation management cases (lower panels). However, the height of the radiative-convective boundary and the stratospheric temperature profiles differ depending on the nature of the climate change, as do the vertical energy fluxes and the heating rates. This is what causes the difference in the entropy production rates.  

Table \ref{tab:sensitivity} lists energy fluxes, temperatures efficiency and entropy production rates for each climate change and system perspective. When the greenhouse gas concentration is increased (second column of Table \ref{tab:sensitivity}) relative to a pre-industrial control scenario (first column), the amount of solar energy absorbed by the system is unchanged, which fixes the emission temperature and with it (approximately) $T_{mat}^{out}$, $T_{tran}^{out}$ and the tropopause temperature $T_{tp}$. Although $F_{tran}$, the total energy transferred, is fixed, the increased optical thickness inhibits heat transfer by radiation, increasing the fraction of energy transferred by material processes ($F_{mat}/F_{tran}$) from $0.47$ to $0.51$. 

\begin{table*}[h!]

\begin{center}
\vspace{.4em}
\begin{tabular}{l || c |c | c | c}

  \multicolumn {1}{c}{ } & \quad Control \quad & \quad GHG Increase \quad & \, Albedo Decrease \, & \quad SRM \quad \\
\hline \hline
 $T_{surf}$ (K)  				& 287.9	&   293.4	&  293.4  &  287.9   \\ 
 $T_{tp}$ (K) 					& 224.1 	& 223.9	&  228.3  &  219.7  \\ 
 $T_{eff}$ (K) 					& 255.1 	& 255.1	&  259.9  &  250.3  \\ 

 \hline
  $F_{mat}$ (W/m$^2$)  			& 111.9 	&  123.5	& 120.6 &  114.6 \\ 
  $T_{mat}^{in}$ (W/m$^2$)  		& 287.9 	&  293.4	& 293.4 &	 287.9 \\ 
  $T_{mat}^{out}$ (W/m$^2$)  		& 269.1	&  270.8	& 274.2 & 265.7 \\ 
  $\eta_{mat}$ (\%)		  		& 6.54 	&  7.71	& 6.54  & 7.71 \\ 
  $\Sigma_{mat}$ (mW/m$^2$K)  	&  27.2	&  35.1	& 28.7   &  33.2   \\ 

 \hline

  $F_{tran}$ (W/m$^2$)  			& 240.0 	&  240.0	& 258.6 & 222.7 \\ 
  $T_{tran}^{in}$ (W/m$^2$)  		& 280.0 	&  284.7	& 285.2 & 279.4 \\ 
  $T_{tran}^{out}$ (W/m$^2$)  		& 257.2	&  257.3	& 262.0 & 252.5 \\ 
  $\eta_{tran}$ (\%)		  		& 8.13	&  9.64	& 8.13   & 9.64 \\
  $\Sigma_{tran}$ (mW/m$^2$K)  	&  75.9	&  89.9	& 80.3   & 85.0  \\ 

\hline

  $F_{planet}$ (W/m$^2$)  		& 342.9 	&  342.9	& 342.9 & 342.9 \\ 
  $T_{planet}^{in}$ (W/m$^2$)  		& 4334 	&  4334 	& 4334 & 4334   \\ 
  $T_{planet}^{out}$ (W/m$^2$)  	& 251.1	& 251.1	& 241.7 & 260.9 \\ 
 $\eta_{planet}$ (\%)		  		& 94.21	& 94.21    & 94.42 & 93.98  \\
  $\Sigma_{planet}$ (mW/m$^2$K)  	&  1286	&  1286	& 1340  & 1235  \\ 

  \hline  
\end{tabular}
\end{center}

\caption{ The values of the entropy-related variables estimated using the radiative-convective model in a unperturbed control climate (first column), under an increased greenhouse effect (second column), a decrease in the global albedo (third column, equivalently an increased solar absorption) and a solar radiation management scenario (fourth column), corresponding to the scenarios plotted in Figure \ref{fig:RCM}. The tropopause temperature, $T_{tp}$ is defined as the temperature minimum, while the effective radiating temperature is a function of the absorbed solar radiation, $T_{eff} = (F_{tran}/\sigma)^{1/4}$.}

\label{tab:sensitivity}
\end{table*}

The increase in the surface temperature while the effective emission temperature and total flux through the system are fixed explains the sharp increase in $\Sigma_{tran}$  and $\eta_{tran}$ by $18\%$. The increased energy flux through the material system results in an even more dramatic $29\%$ increase in the material entropy production rate, with only an $18\%$ increase in $\eta_{mat}$. The planetary entropy production rate is, by contrast, unaltered by this climate change as neither the solar incoming nor the scattered spectra are altered and the shape of the outgoing spectrum can only change minutely given the requirement for a fixed total energy flux in this gray atmosphere.

The story is very different when the same surface temperature increase is achieved by reducing the albedo of the planet (third column of Table \ref{tab:sensitivity}). As the amount of solar radiation absorbed and thermalized rather than scattered increases by $8\%$, the effective emission temperature must necessarily increase by $1.9\%$, as do the $T_{mat}^{out}$, $T_{tran}^{out}$ and $T_{tp}$, while the surface and inflow temperatures increase only marginally more such that the material and transfer efficiencies $\eta$ are conserved. The higher absolute temperatures alone would give a decrease in the entropy production rate by $2\%$, but the increase in energy flux dominates for an overall $6\%$ increase in $\Sigma_{mat}$ and $\Sigma_{tran}$, which is significantly less than in the greenhouse gas case. The decreased scattered fraction causes an increase in $\Sigma_{planet}$ by $4\%$, as thermalized and re-emitted radiation carries significantly more entropy than scattered solar radiation. 

The difference in the entropy responses to surface warming by longwave and shortwave mechanisms reflects the different heating profiles by radiative and material processes in these two cases, even though the surface temperature change is the same. When the albedo change is reversed and combined with the greenhouse gas increase, the surface temperature is restored but not the heating rates nor the entropy production rates, as shown in the fourth column of Table \ref{tab:sensitivity} and in the lower panels of Figure \ref{fig:RCM}. The total absorbed radiation ($F_{tran}$) is reduced by $7\%$ compared to the pre-industrial case, but the fraction of energy transferred by material processes, $F_{mat}/F_{tran}$, is still elevated at $0.51$ due to the greenhouse gases present, which explains the small overall increase in $F_{mat}$. The reduction in absorbed solar radiation also lowers the effective emission temperature of the planet relative to the pre-industrial scenario, and so there is a larger temperature difference between material and transfer influx and outflux temperatures, explaining the higher efficiency. These result in a material entropy production rate that is $22\%$ higher and a transfer entropy production rate that is $12\%$ higher than the control case, although the surface temperature is unaltered. The increased scattering fraction decreases $\Sigma_{planet}$. 

These results are striking for three reasons. Firstly, we find that all three entropy production rates \textit{increase} with solar absorptivity, which differs from the conclusions drawn by \cite{Kato2020} upon regression of observational transfer and material entropy production rates against inter-annual variability of solar absorptivity. Storage of entropy and energy in the oceans in high-absorptivity years could account for the decreases they note, as a top of atmosphere imbalance in net energy flux will result in an imbalanced entropy flux that looks similar to a production. However, the increased greenhouse gas results we find are consistent with previous studies. In a cloud-resolving model, \cite{Singh2016} find an increase in material entropy production rate with an increase in greenhouse gas concentration, as do \cite{Lucarini2010t} in a slab-ocean model. \cite{Bannon2017} also study the response to climate changes in an energy balance model, but without constraints connecting the surface and atmosphere temperatures, so that their results are not directly comparable to ours.

Secondly, the material and transfer efficiencies measured here vary with greenhouse gas concentration but not albedo, as do the ratios $F_{mat}/F_{tran}$ and $\Sigma_{mat}/\Sigma_{tran}$. That the fraction of the transfer entropy production that is due to material processes remains fixed as the solar energy absorbed and temperatures vary suggests that there is physical significance in how internal energy transfer is partitioned between radiative and non-radiative mechanisms.  The increased dominance of  material processes with a greenhouse gas increase is reminiscent of the observed increase in convective mean available potential energy \citep{Gertler2019} and so is not entirely unexpected. Further investigation of this result in models with more precise treatment of humidity appears to be a useful area for further work.

Thirdly, the fact that manipulating the planet's albedo to balance a greenhouse gas change can restore surface temperature while not restoring these entropy metrics underlines that there is useful additional information in these global scalar variables for climate change discussions and decision-making. Entropy production rates have a more direct relationship to the motion and flows in the climate than does global mean surface temperature and, although they are less familiar and so not as easy to interpret, they warrant further exploration as a supplementary diagnostic to advance our understanding of the climate state.

%
%
%
\subsection{Advantages of the transfer perspective in capturing the climate state}
As exemplified by the elevated greenhouse gas case, the planetary entropy production rate is relatively insensitive to climate changes in which the albedo remains fixed, as the outgoing entropy flux is approximately determined by the (unchanged) effective emission temperature and the incoming entropy flux is set by the solar temperature. In fact, an atmosphere-less isothermal rock, with identical solar flux and albedo, will have a similar planetary entropy production rate to a planet with any greenhouse effect. Furthermore, the sun plays an inordinately significant role in the planetary entropy production rate. The value is dominated by the entropy production due to the thermalization of this solar radiation, $\sigma^{atm}_{SW} + \sigma^{surf}_{SW}$ (see Appendix B) and the incoming entropy flux ($\approx \frac{4}{3} F/ T_{sun}$) depends explicitly on the temperature of the sun, although this ought not to influence the climate separately from its role in delivering energy. Taken together, these arguments suggest that the planetary perspective is not a good candidate for studying the climate.

The material entropy production rate is, of the three, the most focused on processes relevant to human experience: it is material processes such as the hydrological cycle or convective motion, and not the internal radiation, which directly feature in the weather we experience. However, not all material processes are equally relevant, and it could be more meaningful to consider them separately; for example, the contribution from frictional dissipation around falling precipitation is twice that from atmospheric motions \citep{Singh2016}, but has a very different significance. The material sub-processes, and even the total material tally, are interdependent portions of a larger system and the energy carried by them can vary because of changes in other parallel processes, such as internal radiation. This makes interpreting changes in the material entropy production rate alone challenging, as they could be due to changes in the proportion of energy transferred rather than in the efficiency or temperature differences. 

The material perspective is also limited in its perception of the greenhouse effect, as this is a fundamentally radiative phenomenon. To demonstrate this, consider the extreme case of a climate without material processes but with a variable greenhouse effect. A surface temperature change caused by an increase in that greenhouse effect would be one that $\Sigma_{mat}$ would be powerless to resolve, material processes being identically null in both cases, even though it would seem to be a climatologically-relevant change. Thus the total material entropy production rate, though meaningful, does not stand out as the most natural climate-summarizing global entropy variable, rather as a variable that describes important sub-processes.

We argue that the transfer entropy production rate is of more compelling value from the perspective of the climate system.  $F_{tran}$, the flux of energy that is absorbed by the planet, is a natural climate variable, as is the temperature difference between where shortwave radiation is absorbed, $T^{in}_{tran}$, and where longwave radiation leaves the planet, $T^{out}_{tran}$. The transfer rate is sensitive to climate changes and reflects all processes that move heat in the planet, without distinguishing between mechanisms. That the fraction of the transfer entropy production by material processes stays fixed as the albedo is changed but varies as the greenhouse gas concentration changes suggests that, in the context of the transfer rate, the material entropy production rate and the entropy production due to internal radiation gain additional significance.

%
%
%

\section{Conclusions}

The total entropy production rate of the Earth is a tantalizing physical concept without a simple interpretation, thanks to the ambiguity in defining the boundary of the climate system with respect to the radiation that feeds it. In this paper we have laid out three options, and with them three entropy production rates that account for the irreversibility of the processes within each of these systems. The planetary perspective includes the entropy production from all radiative and non-radiative processes, whereas the material perspective includes only non-radiative contributions. The transfer perspective separates radiation according to its role within the climate, including only the production due to that which is emitted from and re-absorbed within the system. The exploration of this third option was the particular aim of this paper.
 
 We provide estimates of each entropy production rate in three model climates of varying complexity. The range suggests $\Sigma_{mat} \approx 27$-$48$\mwk, $\Sigma_{tran} \approx 67$-$76$\mwk\ and $\Sigma_{planet} \approx 1279$-$1312$\mwk, which are consistent with the literature. The response of each entropy production rate to climate changes is also explored in a simplified radiative-convective model: the planetary entropy production rate is unchanged by changes in the greenhouse effect and increases with increased shortwave absorption, while the transfer and material entropy production rates increase with surface temperature, but more significantly if that increase is mediated by the greenhouse effect rather than albedo. The fraction of entropy produced by material processes relative to internal radiation is unchanged by albedo changes but increases with greenhouse gas concentration. None of the entropy production rates is restored to pre-industrial levels by solar radiation management following a greenhouse gas increase, if such an intervention restores average surface temperature. 
  
The transfer view of the system has some immediately apparent physical elegance, but work is required to explore its significance further. Although the entropy production rate initially proposed by \cite{Paltridge1975} as a climate-predicting variable is the horizontal component of both the transfer and material perspectives, in the vertical only the extremization of the material entropy production rate has, to our knowledge, been explored (e.g. \cite{Ozawa1997}). Heat transfer by internal radiation in the vertical is of a similar order of magnitude to that by material processes and acts alongside the material processes to transfer heat down-gradient. It is plausible that by considering the sum of these radiative and non-radiative internal heat transfer processes a more coherent view of the climate as a self-optimizing system may emerge.   In fact, the maximum flow theory known as the Constructal Law (discussed in \cite{Reis2014}) appears to suggest a tendency of systems with fixed energy flow, like the climate, to organize to minimize the transfer entropy production rate in particular. More generally, any theory of entropy production extremization must carefully address to which global entropy production variable(s) it applies and why.

 If an entropy-extremization principle were understood in the climate, it might also apply elsewhere. Non-equilibrium quasi-steady systems are common in other areas of complexity, life being one example. In this broader context, the climate can be taken as a convenient, thoroughly-studied and modeled example. 

Fundamental research into the way we interpret and understand the climate has potential societal importance as we wrestle with communication and decision-making based on the digestible knowledge gleaned from complex models. Entropy production rates offer another diagnostic for comparing models to reality and to each other, and for summarizing and tracking climate changes. A predictive theory of entropy generation might also potentially help to constrain climate predictions. This paper's development of the concept of global entropy production is aimed at stimulating further research in this area. 

%
%
%

%
\acknowledgments
With thanks to Helen Brindley, Diane Barnett, Siarhei Barodka, Thomas Bendall, Michael Byrne, Jonathan Gibbins,  Matthew Kasoar and Barnabas Walker for helpful discussions, to Bernhard Mayer for advice on the utilization of Libradtran, to Tyler D. Robinson for example code for the radiative-convective model and to the reviewers for their helpful comments. This work was supported by the EPSRC Mathematics of Planet Earth Centre for Doctoral Training, EP/L016613/1. 
%
%
%

 
\appendix[A]
\appendixtitle{Solution of the EBM}

To make a numerical analysis of the energy balance model, we have used the same values as \cite{Bannon2015}. The script used  in this study is available upon request. 

The EBM set-up is shown in Figure \ref{fig:ebm}. Unlike in a standard EBM, the radiation is separated between that which is internally transferred within the climate and that which is external, i.e. is emitted to space or absorbed from the sun. 

The solar flux is $F_{sun} = 340.7$\wm\ with a fixed albedo of $\alpha = 0.30$ and a solar temperature of $T_{sun} = 5779$\,K. Therefore $F_{SW}^{scat} = \alpha F_{sun} = 102.2$\wm. The atmospheric absorptivity in the shortwave is $\beta = 0.10$, which sets $F_{SW}^{atm} = \beta F_{sun} = 34.1$\wm\ and $F_{SW}^{surf} = (1-\alpha-\beta)F_{sun} = 204.4$\wm, where any surface scattering and subsequent absorption by the atmosphere has been neglected, following \cite{Bannon2015}. The surface-atmosphere material heat fluxes  are represented by a sensible and a latent heat term that transfer heat as a fraction of the incoming solar flux: $F_{sens} = \gamma_{sens} F_{sun} = 17.0$\wm\ and $F_{LH} = \gamma_{LH} F_{sun} = 68.1$\wm\ where $\gamma_{sens} = 0.05$ and  $\gamma_{LH} = 0.2$, approximately following the energy budget of \cite{Wild2014} such that the total $\gamma = 0.25$ matches \cite{Bannon2015}. The emissivity of the atmosphere is $\epsilon = 0.95$ and so $F_{LW}^{atm} = \epsilon \sigma T_{atm}^4$, while the surface is a black body such that $F_{surf} = \sigma T_{surf}^4$. Requiring energy balance for the surface and atmosphere gives the temperatures $T_{surf} = 278.4$\,K and $T_{atm} = 253.2$\,K. From these, the values of the longwave energy fluxes can be found: $F_{intrad} = \epsilon \sigma T_{surf}^4 - \epsilon \sigma T_{atm}^4 = 102.2$\wm, $F_{LW}^{surf} = (1-\epsilon) \sigma T_{surf}^4 = 17.0$\wm and $F_{LW}^{atm} = \epsilon T_{atm}^4 = 221.5$ W/m$^2$, as demonstrated in \cite{Bannon2015}. (It is coincidental that in this example the internal radiative heat transfer is the same value as the scattered energy).

%
%
%

\appendix[B]
\appendixtitle{Calculation of EPRs by the direct method}

Approximating to black body behavior, the absorption of radiation results in an entropy production of the form of a difference between the entropy of the heat in the material and the entropy in the radiation:
\begin{equation}
\sigma(\text{absorb}) = \frac{F}{T_{mat}} - \frac{4}{3} \frac{F}{T_{source}}
\end{equation}
which is reversed for emission. For black body emission, the temperature of the relevant material $T_{mat}$ will also be the source temperature $T_{source}$.

The entropy production due to an internal heat transfer is:
\begin{equation}
\sigma(\text{internal heat transfer}) = F \left( \frac{1}{T_{cold}} - \frac{1}{T_{hot}}\right)
\end{equation}
which applies both to the material and to the internal radiation terms. The spectral character of the internal radiation need not be accounted for in the entropy production term because both the creation and destruction of those photons happen within the planetary boundaries; only the resultant heating is relevant. 

These principles can now be applied to the EBM in order to calculate directly the three total entropy production rates, beginning by calculating the contributions from each process.

The entropy produced upon scattering is generally a function of the change in directional intensity of the radiation. Following \cite{Bannon2015} and \cite{Wu2010},  $J_{scat} = 110.0$\,\mwk\ so that the production rate is: 
\begin{equation*}
\sigma^{scat}_{SW} = J_{scat} - \frac{4}{3} \frac{F_{scat}}{T_{sun}}  = 86.4\ \text{\mwk}.
\end{equation*}

The thermalization of solar radiation in the atmosphere results in an entropy production of:
\begin{equation*}
\sigma^{atm}_{SW} = F_{SW}^{atm} \left( \frac{1}{T_{atm}} - \frac{4}{3 T_{sun}}\right) = 126.7 \ \text{\mwk}
\end{equation*}
\noindent and similarly, the thermalization of solar radiation at the surface results in:
\begin{equation*}
\sigma^{surf}_{SW} = F_{SW}^{surf} \left( \frac{1}{T_{surf}} - \frac{4}{3 T_{sun}}\right) = 687.1\ \text{\mwk}.
\end{equation*}

The material transport of heat from the surface to the atmosphere causes much smaller entropy productions, proportional to the reciprocal temperature difference:
\begin{equation*}
\sigma_{sens} = F_{sens} \left( \frac{1}{T_{atm}} - \frac{1}{T_{surf}}\right) = 6.1\ \text{\mwk}
\end{equation*}

\begin{equation*}
\sigma_{LH} = F_{LH} \left( \frac{1}{T_{atm}} - \frac{1}{T_{surf}}\right) = 24.4\ \text{\mwk}
\end{equation*}
\noindent as does the net transport of heat from the surface to the atmosphere via internal radiation 
\begin{equation*}
\sigma_{intrad} = F_{intrad} \left( \frac{1}{T_{atm}} - \frac{1}{T_{surf}}\right) = 36.5\ \text{\mwk}.
\end{equation*}

Emission of longwave radiation from the surface results in an entropy production because the radiation carries more entropy than the cooled matter loses. For the emission from the surface:
\begin{equation*}
\sigma^{surf}_{LW} = F_{LW}^{surf} \left( \frac{4}{3}\frac{1}{T_{surf}} - \frac{1}{T_{surf}}\right) = 20.4\ \text{\mwk}
\end{equation*}
\noindent and for the emission from the atmosphere:
\begin{equation*}
\sigma^{atm}_{LW} = F_{LW}^{atm} \left( \frac{4}{3}\frac{1}{T_{atm}} - \frac{1}{T_{atm}}\right) = 291.5\ \text{\mwk}.
\end{equation*}

From these constituent budget terms, the three entropy productions rates can be calculated directly. The planetary entropy production is the total from all processes:
\begin{multline}
\Sigma_{planet} = \sigma^{scat}_{SW} + \sigma^{atm}_{SW} + \sigma^{surf}_{SW} + \sigma_{LH} + \sigma_{sens}\\ + \sigma_{intrad} + \sigma^{surf}_{LW} + \sigma^{atm}_{LW} = 1279.1\ \text{\mwk}
\label{eq:PEPRcalc}
\end{multline}
which is in agreement with \cite{Bannon2015} for CV1.

The material entropy production rate excludes all radiative processes:
\begin{equation}
\Sigma_{mat} = \sigma_{LH} + \sigma_{sens} = 30.4 \ \text{\mwk}.
\label{eq:MEPRcalc}
\end{equation}

The transfer entropy production rate includes only the processes that involve energy exchange between parts of the material system, by both internal radiation and material processes:
\begin{equation}
\Sigma_{tran} = \sigma_{LH} + \sigma_{sens} + \sigma_{intrad} = 67.0 \ \text{\mwk}.
\label{eq:teprdirect}
\end{equation}
%

%
%
%

\appendix[C]
\appendixtitle{Comparison with the indirect method}

These values can also be calculated via the indirect method, using the relationship $\Sigma = J_{out} - J_{in}$ under the assumption of steady state.

The sun as a black body at temperature $T_{sun}$ carries entropy towards the earth of 
\begin{equation}
J^{in}_{planet} = \frac{4}{3} \frac{F_{sun}}{T_s} = 78.6\ \text{\mwk}
\end{equation}
\noindent and the outgoing radiation carries entropy according to its emission temperature
\begin{equation}
J^{out}_{planet}  = J_{scat} + \frac{4}{3}\frac{F_{LW}^{surf}}{T_{surf}}  + \frac{4}{3}\frac{F_{LW}^{atm}}{T_{atm}} = 1357.7\ \text{\mwk}
\end{equation} 
such that the difference is the planetary entropy production rate of Equation \ref{eq:PEPRcalc}. The entropy fluxes can also be used to calculate representative temperatures via the relationship $T = F/J$ where $F_{planet} = 341$\wm. Then $T^{in}_{planet} = 4334.2$\,K and $T^{out}_{planet} = 250.9$\,K.

For the transfer entropy production rate, the flux of entropy into the sub-system is the entropy change of the material upon absorption of solar radiation:
\begin{equation}
J^{in}_{tran} = \frac{F_{SW}^{atm}}{T_{atm}}  + \frac{F_{SW}^{surf}}{T_{surf}}= 868.8\ \text{\mwk}.
\end{equation}
\noindent The flux of entropy out is the change of entropy of the material that is due to cooling to space:
\begin{equation}
J^{out}_{tran} =  \frac{F_{LW}^{surf}}{T_{surf}} + \frac{F_{LW}^{atm}}{T_{atm}} =935.8\ \text{\mwk}.
\end{equation} 
The difference, $67.0$\mwk, is identical to the result of the direct calculation using Equation \ref{eq:teprdirect} above. The flux of energy in this case is $F_{tran} = F_{SW}^{atm} +F_{SW}^{surf} = 239$\wm, which results in representative temperatures $T^{in}_{tran} = 274.5$\,K and $T^{out}_{tran} = 254.9$\,K. The out-flux temperature is close to the effective radiating temperature of the planet $T_{eff}^* = \left((1-\alpha) F_{sun}/\sigma \right)^{1/4} = 254.7$\,K, as expected.

For the material entropy production rate there is a choice about how to separate flux into and out of the system. We take the approach of calculating the net radiative heating rate at each point and separating the system into areas of net radiative heating (the surface) and radiative cooling (the atmosphere):
\begin{equation}
J^{in}_{mat} = \frac{F_{SW}^{surf} - F_{LW}^{surf} - F_{intrad}}{T_{surf}}  = 306.0 \ \text{\mwk}
\end{equation}
\begin{equation}
J^{out}_{mat} =  \frac{F_{SW}^{atm} + F_{intrad} - F_{LW}^{atm}}{T_{atm}} = 336.4 \ \text{\mwk}.
\end{equation} 
Again the difference in entropy flux agrees with the production calculated above (Equation \ref{eq:MEPRcalc}).  The flux through this version of the system is $F_{mat} = 85.18$\wm\ and the temperatures are accordingly $T^{in}_{mat} = 278.4$\,K and $T^{out}_{mat} = 253.2$\,K. These values are summarized in Table \ref{tab:values}.

%
%
%

\appendix[D]
\appendixtitle{Analytic radiative-convective model definition}

The analytic radiative-convective model used here is the one described in \cite{Tolento2019}, developed from earlier work in \cite{Robinson2012, Robinson2014}. It is designed to be simple enough to solve analytically and versatile enough to fit a range of planetary atmospheres. Here we describe its application to Earth in particular, using the parameter values from \cite{Tolento2019}. Profiles of the temperature, energy fluxes and heating rates calculated from this model are shown in Figure \ref{fig:RCM}, solid lines. Our script to run this model is available by request.

The vertical coordinate of the model, $\tau$, is the gray thermal optical depth and the atmosphere is split into two portions: a stratospheric part, which is in radiative balance, and a tropospheric part where convection occurs.  In the lower portion, the thermal structure is given by a modified adiabat up to the radiative-convective boundary at $\tau_{rc}$:

\begin{equation}
T = T_0 \left( \frac{\tau}{\tau_0} \right) ^{\beta/n}
\end{equation}

where $T_0$ is the reference temperature at the $\tau_0$, the surface of Earth, and $n=2$ establishes the relationship between optical depth and pressure via $\tau/\tau_0 = (p/p_0)^n$, with $p_0 = 1$\,atm. The parameter $\beta = a (\gamma-1)/\gamma$ extends the dry adiabatic lapse rate (in terms of the ratio of specific heats $\gamma = 1.4$) to account for latent heat release via the rescaling by $a = 0.6$, which establishes the slope of the temperature profile shown in the first panel of Figure \ref{fig:RCM}.

The net solar radiative flux is given by two shortwave channels, which attenuate as a function of the thermal optical depth:
\begin{equation}
F^{\odot} = \alpha \left( F_1^\odot e^{-k_1 \tau} + F_2^\odot e^{-k_2 \tau}\right)
\end{equation}
with $F_1^\odot = 10$\wm, $k_1 = 90$, $F_2^\odot = 333$\wm, $k_2 = 0.16$ and $\alpha$ the top of atmosphere albedo.  The impact of these two channels can be seen in the third panel of Figure \ref{fig:RCM}: the stratospheric peak in the solar heating rate is accomplished by $F_1^\odot$, while the tropospheric and surface solar absorption is due to $F_2^\odot$. The entropy of the scattered flux is approximated as in \cite{Stephens1993, Bannon2015, Wu2010} by assuming isotropic (Lambertian) scattering, so that the $J_{scat} = \frac{4}{3} \sigma T_{sun}^3 \chi(u_L)$ where $ \chi(u) \approx u(-0.2776 \ln{u} + 0.9651)$ and $u_L =  \alpha \Omega_{sun}/4\pi$, where $\Omega_{sun} = 6.77 \times 10^{-5}$ st and $T_{sun}=5779$\,K.  

The radiative transfer in both regions is given by the gray two-stream Schwarzschild equations:
\begin{align}
\frac{dF^+}{d\tau} &= D \left( F^+ - \sigma T^4 \right) \\
\frac{dF^-}{d\tau} &= -D \left( F^- - \sigma T^4 \right) 
\end{align}
where $D = 1.66$ is the diffusivity factor and $F^+$ and $F^-$ are the upwelling and downwelling longwave radiative fluxes. Integrating these two equations and plugging in the solar flux and the convective temperature profile provides, upon further manipulation, expressions for the upwelling thermal flux and the temperature in both the convective and non-convective region, in terms of incomplete gamma functions that can be handled numerically (for derivation, see \cite{Robinson2012, Tolento2019}). The two constraints -- that the upwelling radiative flux and temperature be continuous across the radiative-convective boundary -- then allows the model to be solved for two free parameters (for example, $\tau_{rc}$ and $T_0$) by standard root-finding methods. We take the total column optical thickness to be $1.96$ and the global albedo to be $0.3$ in the unperturbed case, resulting in a surface temperature of $287.9$\,K. The convective heat flux is found as the difference between the the net upwards LW radiative flux ($F_{net}^{LW} = F^+ - F^-$) and downwelling net solar flux, $F_{conv}(\tau) = F_{net}^\odot(\tau) - F_{net}^{LW}(\tau)$.

 \bibliographystyle{ametsoc2014}
 \bibliography{TEPRBibliography}



\end{document}